\documentclass[
 superscriptaddress,
reprint,
showpacs,
 bibnotes,
 showkeys,
 amsmath,amssymb,
 aps,
]{revtex4-1}
\usepackage{slashbox,pict2e}
\usepackage{graphicx}
\usepackage{amsmath}
\usepackage{dcolumn}
\usepackage{bm}
\usepackage{braket} 
\usepackage{bbold}
\usepackage[
 colorlinks=true,
 urlcolor=blue,
 citecolor=blue,
 linkcolor=blue,
 bookmarks=false,
 pdfstartview={FitH},
]{hyperref}

\def\EFA{EuFe$_2$As$_2$}
\def\Eg{$E_g$}
\def\Bx{$B_{2g}$}
\def\By{$B_{3g}$}
\def\cm{cm$^{-1}$}
\def\Ts{$T_S$}

\begin{document}

\title{Stress-induced nematicity in \EFA\, studied by Raman spectroscopy}
\author{W.-L.~Zhang}
\email{wlzhang@iphy.ac.cn}
\affiliation{Beijing National Laboratory for Condensed Matter Physics and Institute of Physics, Chinese Academy of Sciences, Beijing, 100190, China} 
\affiliation{Department of Physics $\&$ Astronomy, Rutgers University, Piscataway, New Jersey 08854, USA} 
\author{Athena~S.~Sefat}
\affiliation{Materials Science and Technology Division, Oak Ridge National Laboratory, Oak Ridge, Tennessee 37831-6114, USA}  
\author{H.~Ding}
\affiliation{Beijing National Laboratory for Condensed Matter Physics and Institute of Physics, Chinese Academy of Sciences, Beijing, 100190, China} 
\affiliation{Collaborative Innovation Center of Quantum Matter, Beijing, China} 
\author{P.~Richard}
\affiliation{Beijing National Laboratory for Condensed Matter Physics and Institute of Physics, Chinese Academy of Sciences, Beijing, 100190, China} 
\affiliation{Collaborative Innovation Center of Quantum Matter, Beijing, China}
\author{G.~Blumberg}
\email{girsh@physics.rutgers.edu}
\affiliation{Department of Physics $\&$ Astronomy, Rutgers University, Piscataway, New Jersey 08854, USA}
\affiliation{National Institute of Chemical Physics and Biophysics, Akadeemia tee 23, 12618 Tallinn, Estonia}

\date{\today}


\begin{abstract}

We use polarized Raman scattering to study the structural phase transition in \EFA, the parent compound of the 122-ferropnictide superconductors. The in-plane lattice anisotropy is characterized by measurements of the side surface with different strains induced by different preparation methods. We show that while a fine surface polishing leaves the samples free of residual internal strain, in which case the onset of the $C_4$ symmetry breaking is observed at the nominal structural phase transition temperature \Ts, cutting the side surface induces a permanent four-fold rotational symmetry breaking spanning tens of degrees above \Ts.

\end{abstract}


\maketitle

The 122-ferropnictide superconductors go through a structural phase transition at a temperature \Ts\,that coincides with or precedes a magnetic phase transition at a temperature $T_N$~\cite{Stewart_RMP2011}. In most of the parent and under-doped ferropnictides, measurements of electronic anisotropy are reported below \Ts~\cite{Fisher_Review2011,Terashima_PRL2011,Uchida_PRL2012,LuDai_Science2014}. Above \Ts, unexpected anisotropy is found to be persistent in experiments performed under uniaxial strain~\cite{Fisher_Science2010,Fisher_Review2011,YiM_PNAS2011,Orenstein_1507} or magnetic field~\cite{Matsuda_Nature2012}, which implies a nematic phase transition at a temperature $T^*>T_S$. 
However, other spectroscopic methods claim the absence of such nematic transition. Instead, dynamic nematic fluctuations are already present at room temperature and accumulate gradually upon cooling~\cite{ning_PRL2010,Yoshizawa_JPSJ2012,Goto_JPSJ2011,bohmer_PRL2014,Gallais_PRL2013,YXYang_JPS2014,Hackl_arXiv1507,Verner_PRB2016,Zhang_QEP,Massat_1603}.
By removing twin domains, uniaxial strain breaks the four-fold rotational symmetry $C_4$~\cite{Dhital_PRL2012}, transforming the structural phase transition into a crossover spanning a measurable temperature range above \Ts\,\cite{Wilson_PRB2009,LiYuan_PRL2015}. Consequently, the nematic phase transition above \Ts\, is not universally accepted.

In this paper we study the temperature evolution of stress-induced nematicity above and below the structural phase transition in \EFA, the parent compound of the 122-ferropnictide superconductors. We observe the splitting of the doubly degenerate Fe-As in-plane displacement phonon mode when the lattice $C_4$ symmetry is broken. We measure this splitting below the structural transition temperature and demonstrate that the splitting is directly proportional to the lattice nematic order parameter. We show that stress occurring during the sample preparation induces permanent $C_4$-symmetry breaking strain fields that are distinct from dynamic nematic fluctuations above the tetragonal to orthorhombic structural transition. 

The \EFA\, single crystals (with \Ts\,=\,175~K) used in this Raman study were synthesized by a Fe-As flux method~\cite{Sefat_PRL2008}. We performed Raman scattering from the $ac$ surface prepared by a razor blade cut or fine sandpaper polishing at room temperature. The mechanical polishing has been performed with aluminum oxide sandpapers of several sizes down to 0.1~$\mu$m. We used high purity methanol as lubricant. For each sandpaper, the polished thickness on the sample was precisely controlled to be more than three times the grit size. To minimize strain, we used wax to glue the sample and the adhering point was far away from the measured surface. 

We performed the Raman measurements in a quasi-back-scattering optical setup. The 647 and 752 nm wavelength Kr$^+$ laser beams were focused to a 50$\times$100 $\mu$m spot on the $ac$ surface of the \EFA\, samples. The incident laser power was kept smaller than 10 mW with an estimated 5~K for the laser heating. The laser heating was further verified by the appearance of bright stripes on the $ab$ surface at \Ts~\cite{Prozorov_PRB2009,Hackl_arXiv1507} using the same incident power. Temperature-dependent measurements from 30 to 300 K were preformed in a He gas cooled cryostat. The Raman signal was collected and analyzed by a triple spectrometer and a liquid N$_2$ cooled CCD. The Raman susceptibility $\chi^{\prime\prime}(\omega)$ was calculated using $I(\omega)$=$(1+n(T))\chi^{\prime\prime}(\omega)$, where $I(\omega)$ is the scattering intensity corrected for the system background and the system optical response, and $n(T)$ is the Bose factor.

\begin{figure}[!t]
\begin{center}
\includegraphics[width=\columnwidth]{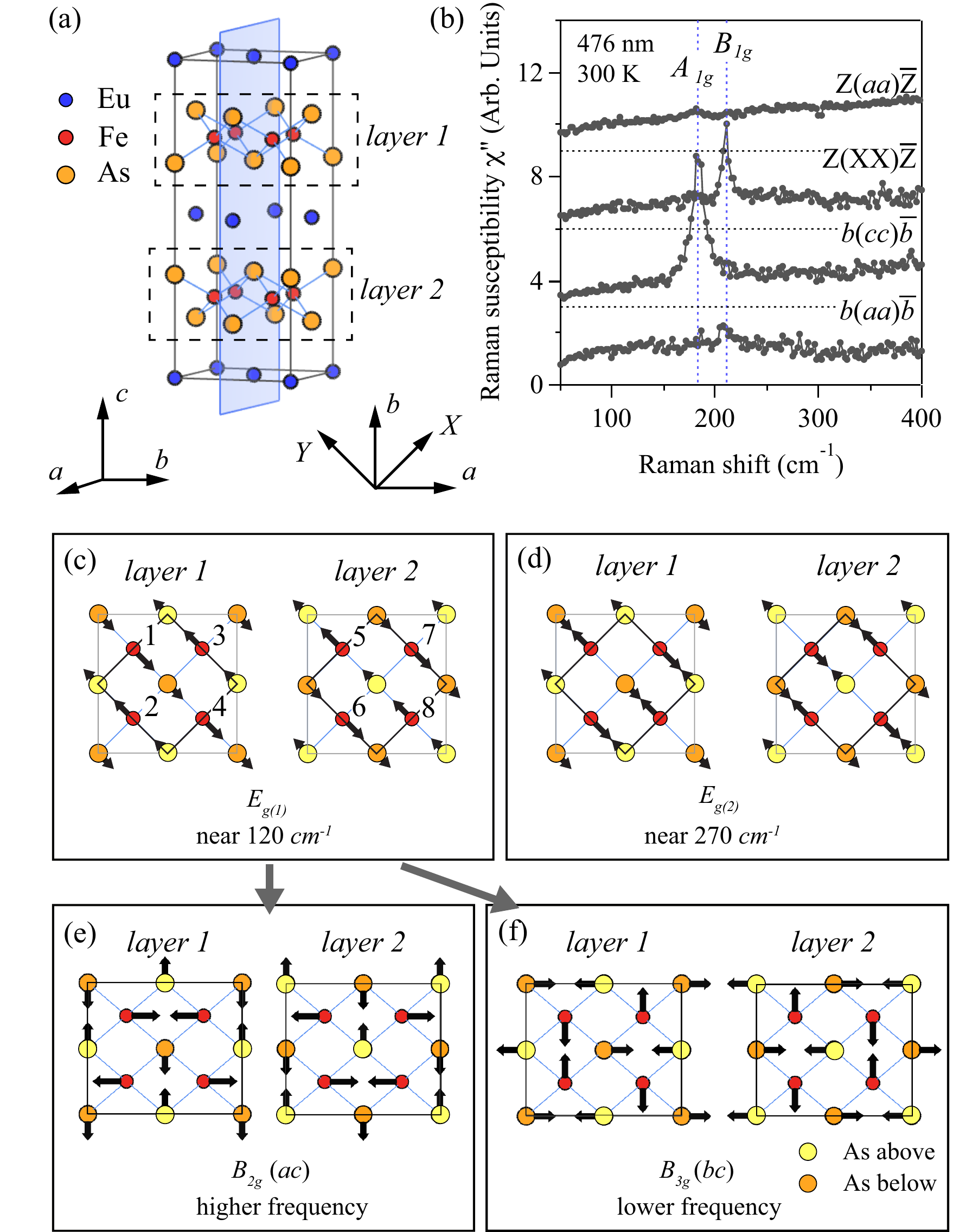}
\end{center}
\caption{\label{Fig: 1} (Color online) (a) The orientation of the measured surface (blue rectangle) and the definition of the axis. The light polarization $c$ is defined along the longest axis of the crystal and the polarizations $a$ and $b$ are the longer and shorter nearest Fe-Fe directions, respectively. The X and Y axes are at 45 deg from $a$ and $b$. (b) Raman susceptibility $\chi^{\prime\prime}$ from the side surface and the $ab$ surface. (c) and (d) Atomic displacements of the two \Eg modes in the high-temperature tetragonal phase.  (e) and (f) Atomic displacements of the \Bx\, and \By\, phonons in the low-temperature phase derived from the lower energy branch of the two degenerate \Eg\, phonons in the high-temperature phase shown in (c). }
\end{figure}

The crystal structure of the 122 ferropnictides in the high-temperature tetragonal phase belongs to space group $I4/mmm$ (point group $D_{4h}$). 
The corresponding energy and atomic displacements of the Raman active phonons (1$A_{1g}$+1$B_{1g}$+2\Eg) at room temperature have been reported previously~\cite{Litvinchuk_PRB2008}. Below \Ts\,the crystal structure belongs to space group $Fmmm$ (point group $D_{2h}$) and the breakdown of the $C_4$ symmetry splits the degenerate \Eg\,mode into \Bx\,and \By. In the four-Fe unit cell basis (X-Y coordinates shown in Fig.~\ref{Fig: 1}, which is rotated for 45 deg from the two-Fe unit cell basis), the Raman tensor of the \Eg\,symmetry in the high-temperature phase and that of the \Bx\,and \By\,symmetries in the low-temperature phase are~\cite{Bilbao_1}:

\[ R_{E_g} = \left( \begin{array}{ccc}
0 & 0 & e \\
0 & 0 & e\\
e & e & 0 \end{array} \right), \,\,
R_{E_g} = \left( \begin{array}{ccc}
0 & 0 &  -e\\
0 & 0 & e\\
-e & e & 0 \end{array} \right),\]

\[ R_{B_{2g}} = \left( \begin{array}{ccc}
0 & 0 & e^\prime \\
0 & 0 & 0\\
e^\prime & 0 & 0 \end{array} \right),\,\,
R_{B_{3g}} = \left( \begin{array}{ccc}
0 & 0 & 0\\
0 & 0 & f^\prime\\
0& f^\prime & 0 \end{array} \right).\]
%


For the 122 family of iron pnictides, with the body centered lattice, in the high-temperature phase, the long-wavelength lattice displacement at the Brillouin-zone center $\Gamma$ point are defined by the translational basis vectors of the primitive cell, which require the Fe ion labeled with 1 (Fe-1) to be in phase with Fe-4, Fe-6, and Fe-7 (Fig.~\ref{Fig: 1}(c)). Similarly, Fe-2, Fe-3, Fe-5, and Fe-8 are in phase. For \Eg(\Bx/\By) symmetry, Fe-1 and Fe-5 are anti-phase as required by the horizontal mirror operation ($\sigma_h$). Hence, Fe-1 and Fe-2 vibrate in anti-phase. The same applies for the As sites. In Figs.~\ref{Fig: 1}(c-d) we illustrate the atomic displacements of the two \Eg\,phonons in the high-temperature phase (Figs.~\ref{Fig: 1}(c) and (d)), and \Bx\,and \By\,phonons derived from the lower-energy branch in the low-temperature phase (Figs.~\ref{Fig: 1}(e) and (f)). Our analysis is consistent with Refs.~\cite{Zbiri_PRB2009,LiYuan_PRL2015}. Here we only consider the symmetry operations, whereas the relative length of the arrows on the Fe and As do not contain information on the vibration amplitude.

The \Bx\, mode is active for $ac$ polarization, whereas the \By\, mode is active for $bc$ polarization. Unlike many other symmetry-sensitive probes that require external uniaxial field to eliminate the average effect from twin domains, the \Bx\, and \By\, modes can both be detected when there are naturally formed twin domains, which allows the measurement of the lattice anisotropy in a free standing sample.

\begin{figure}[!t]
\begin{center}
\includegraphics[width=\columnwidth]{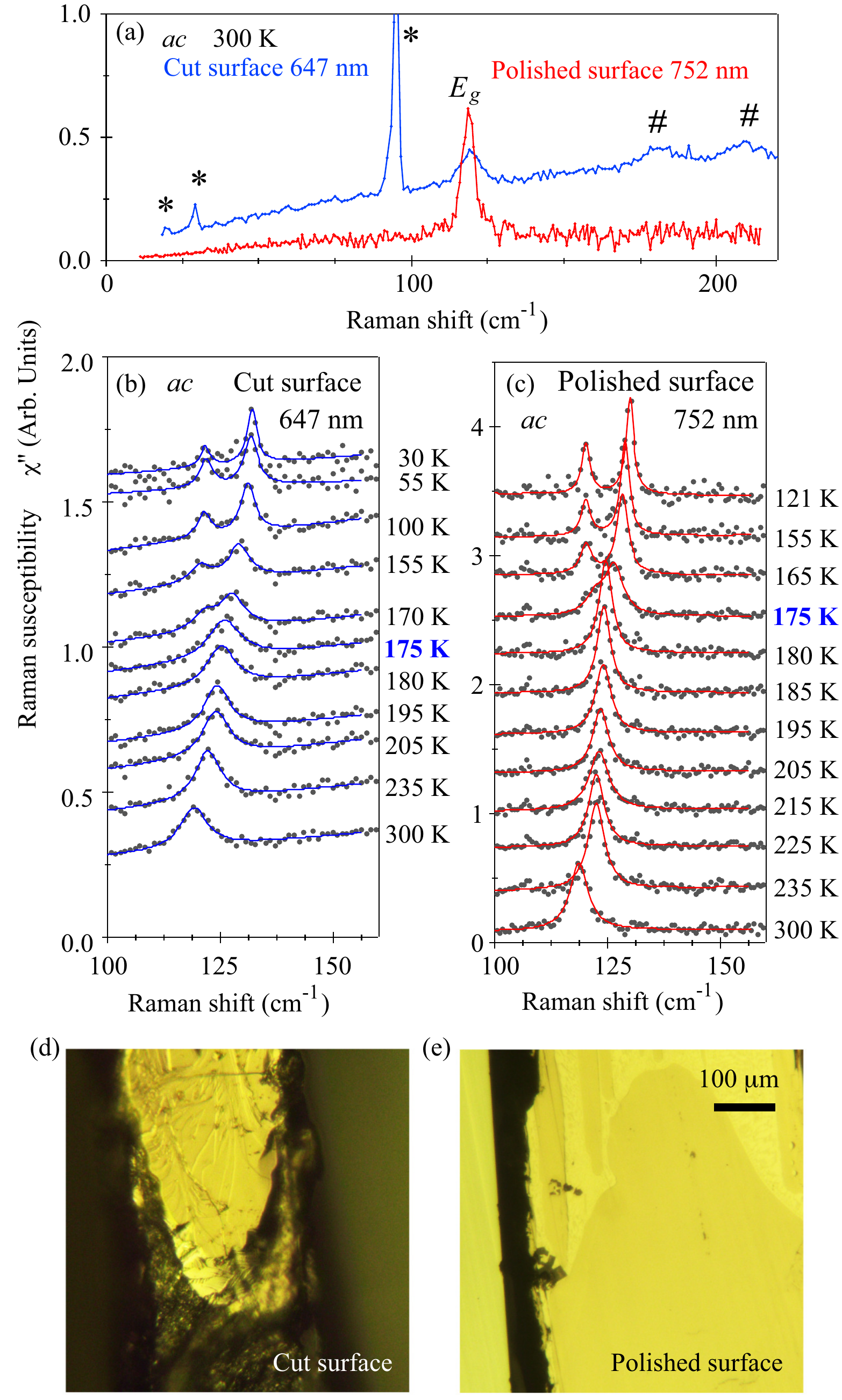}
\end{center}
 \caption{\label{Fig: 2} (Color online) (a) Raman susceptibility measured with the $ac$ polarization configuration at different temperatures on a surface made by a razor blade cut (blue curve) and a surface made by sandpaper polishing (red curve). The three peaks at 19.4, 29.2 and 95.4~\cm\, marked with stars are laser plasma lines. The two peaks at 181 and 208 \cm\, marked with pound signs are the $A_{1g}$ and $B_{1g}$ phonon modes. (b) and (c) Temperature dependent phonon spectra measured with the $ac$ polarization. The blue and red curves are the fits to Lorentz functions.  (d) and (e) Microscopic images of the cut surface and the polished surface, respectively. The reference for the space scaling of the images is given in (e).}
\end{figure}

The orientation of the side surface we obtain is shown by the blue rectangle in Fig.~\ref{Fig: 1}(a).  We further justify it by selection rules of the $B_{1g}$ phonon for different polarization configurations. According to the selection rules, $I_{\mathrm{XX}} =A_{1g}+B_{1g}$, $I_{aa} =A_{1g}+B_{2g}$, and $I_{cc} = A_{1g}$.  In Fig.~\ref{Fig: 1}(b) we show the spectra: $aa$ polarization configurations measured from the $ab$ surface,  XX measured from the $ab$ surface, $cc$ measured from the side surface, and parallel polarizations in the $ab$ plane measured from the side surface from top to bottom. In the last spectrum, the $B_{1g}$ phonon is absent, which indicates that the polarization configuration is $aa$, instead of XX. This confirms that the measured surface is the $ac$ surface.

Fig.~\ref{Fig: 2}(a) shows Raman spectra measured with cross polarizations ($ac$) at room temperature. From a surface obtained by razor blade cut (blue curve), the spectra show six peaks between 18 and 220~\cm. The peak around 120~\cm\,is the lower-energy branch of the \Eg\,mode~\cite{Chauviere_PRB2009}. The three sharp peaks at 19.4, 29.2, and 95.4~\cm\,marked with stars, are the laser plasma lines. 
By comparison with previous phonon measurements on the 122 ferropnictide materials~\cite{Litvinchuk_PRB2008},
we assign the two modes at 183 and 214~\cm\,marked with pound signs to the $A_{1g}$ and $B_{1g}$ phonons, respectively. These two modes should only be Raman active for the in-plane and ZZ polarization configurations.  
The observation of $A_{1g}$ and $B_{1g}$ phonons in the $ac$ polarization configuration suggests the measured surface is bent or contains fragment pieces induced by the cut with the razor blade.

In order to improve the surface quality, we polished the cutting surface to optical flatness (Fig.~\ref{Fig: 2}(e)). From the polished surface (red curve), the scattering background and the laser plasma lines are greatly suppressed. The Raman scattering signals from other symmetry channels are also removed.

In Figs.~\ref{Fig: 2}(b) and (c) we show Raman spectra measured with $ac$ polarization from the two surfaces at different temperatures above and below $T_S$. A clear splitting of the \Eg\,mode is observed below 175 K. 

In order to extract further information about the structural transition, we fit the \Eg\, mode for $T>T_S$ using a single Lorentz function and a linear background:
 
\begin{equation}
\chi_{ac}^{\prime\prime}(\omega, T) = Lor (\omega, \omega_0,\gamma_0,A_0) + a\omega+b,
\end{equation}

\noindent while for $T\leq T_S$ we add another Lorentz term to account for the splitting of \Eg\, into $B_{2g}+B_{3g}$:

\begin{eqnarray}
\chi_{ac}^{\prime\prime}(\omega, T) &=&Lor (\omega, \omega_1,\gamma_1,A_1)\notag\\
 &+& Lor (\omega, \omega_2,\gamma_2,A_2) + a\omega+b,
\end{eqnarray}

\begin{figure}[!t]
\begin{center}
\includegraphics[width=\columnwidth]{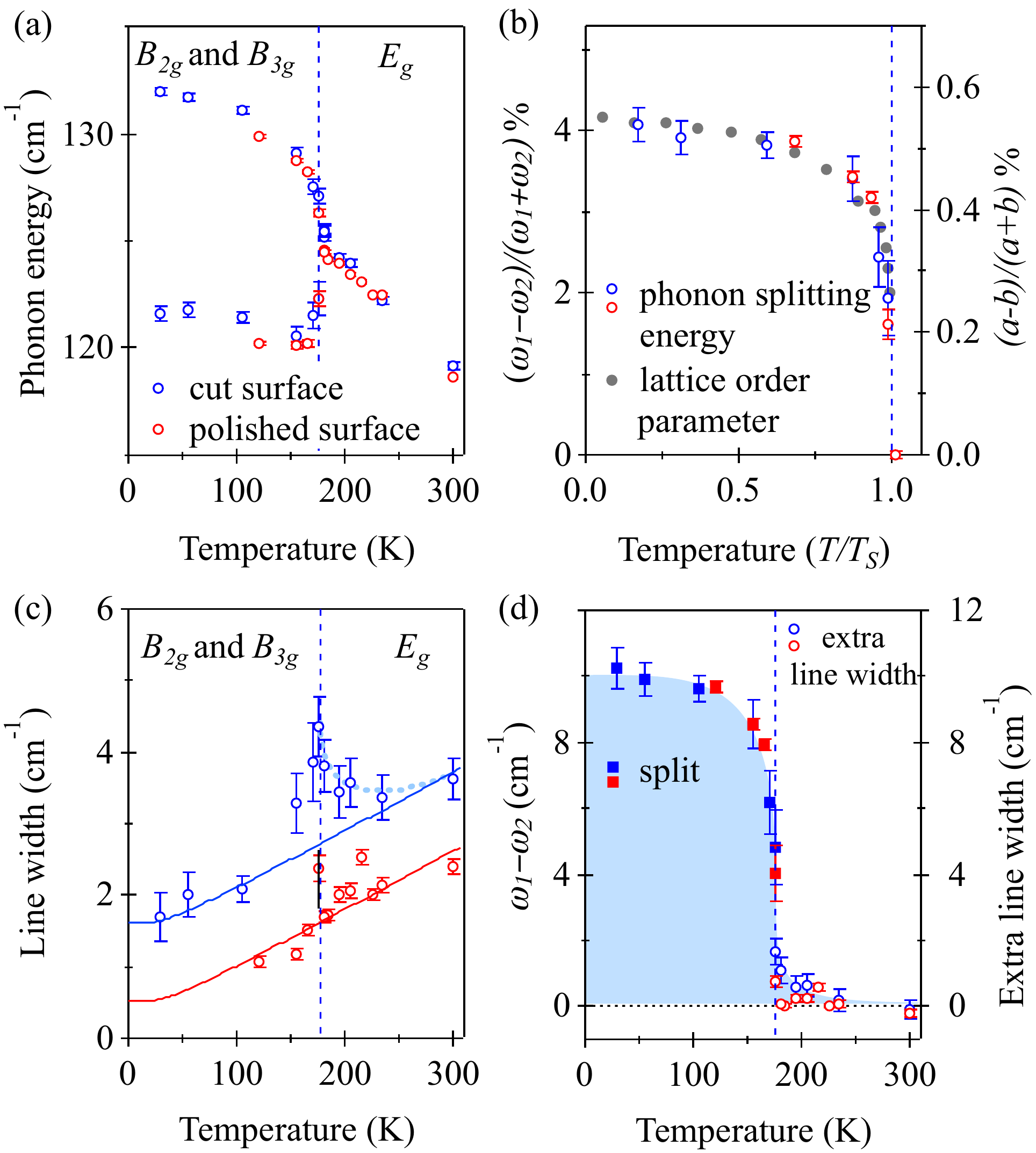}
\end{center}
 \caption{\label{Fig: 3} (Color online)  (a) Phonon energy from the cut surface (blue) and polished surface (red) by fitting to the Lorentz function using Eqs. (1) and (2) above and below \Ts, respectively. The dashed vertical line indicates \Ts. (b) Phonon energy anisotropy and lattice anisotropy order parameter as a function of temperature. The temperature is scaled by \Ts. (c) Phonon linewidth from the two different surfaces. The solid curves are guide lines to the eye. (d) Comparison of the \Eg\, mode extra broadening and \Bx/\By\, mode splitting for the two surfaces. The vertical error bars are from the fitting error. The temperature error is $\pm$ 5~K.}
\end{figure}

\noindent In these expressions $Lor\,(\omega, \omega_i,\gamma_i,A_i) = A_i[(\omega-\omega_i)^2+\gamma_i^2]^{-1}$ is the phonon response, $\omega_i$ is the central energy, $\gamma_i$ is the phonon damping, and $a\omega+b$ is a linear approximation of the background. The fitting curves are displayed in Fig.~\ref{Fig: 2} and the parameters obtained from the fits are plotted in Fig.~\ref{Fig: 3}. The energy of the \Eg\, phonon, as well as its \Bx\, and \By\, components in the low-temperature phase, are almost identical for the two differently treated surfaces (Fig.~\ref{Fig: 3}(a)). As shown in Fig.~\ref{Fig: 3}(b), the phonon energy anisotropy $\frac{\omega_1-\omega_2}{{\omega_1+\omega_2}}$ can be linearly scaled with the lattice orthorhombic order parameter $\frac{a-b}{a+b}$ in the low-temperature phase~\cite{Tegel_JPCM2008}.

Unlike the phonon energies, the phonon linewidths are strongly dependent on the surface measured. The linewidth of the cut surface $\gamma_{cut}$ has an overall broadening of about 1.1 \cm\,compared to the linewidth of the polished surface $\gamma_{polished}$ (Fig.~\ref{Fig: 3}(c)). We attribute the overall broadening to the inhomogeneity of the cut surface. Surprisingly, while the decrease of $\gamma_{polished}$ can be fitted with the expression of the anharmonic decay~\cite{Cardona_PRB1984},
\begin{equation}
\gamma_{ph}(T)=\Gamma_0+\Gamma_1(1+\frac{2}{e^{h\omega_0/2k_BT}-1}),
\end{equation}
with $\omega_0=129.2\pm0.6$ \cm\,obtained from $\omega_{polished}$ in the high-temperature phase, $\Gamma_0=0.1\pm0.3$ \cm, and $\Gamma_1=0.4\pm0.1$ \cm, $\gamma_{cut}$ shows an unusual extra broadening near \Ts. The lowest linewidth from the cut surface in the high-temperature phase is at 235 K, which is 60 K above \Ts. 
Interestingly, a similar linewidth broadening is also reported in the \Eg\, mode and the (0,2,0) or (2,0,0) Bragg peaks in BaFe$_2$As$_2$ under uniaxial strain~\cite{LiYuan_PRL2015,Chen_PRB2016}. Here we stress that the absence of extra broadening for the \Eg\, mode in the polished sample indicates that the nematic fluctuations are frozen or negligible above \Ts, in contrast to previous reports~\cite{Fisher_Science2010,Fisher_Review2011,YiM_PNAS2011,Orenstein_1507,Matsuda_Nature2012}. As a corollary, the extra broadening in the cut sample thus suggests that the $C_4$ symmetry is broken above \Ts\, by internal strain rather than by intrinsic dynamic nematic fluctuations. In addition, the difference in the behaviors observed for the two samples reveals the sensitivity of the Fe-based superconductors to sample preparation. 

Since the \Bx\, and \By\, phonons cannot be distinguished individually, we conjecture that (1) the strain-induced anisotropy is small compared to the linewidth and the energy resolution and (2) the distribution of the strain is inhomogeneous and it results in a continuous energy splitting. Above \Ts\, the \Bx\, and \By\, phonon splitting energy can be approximated by the extra line width broadening. In Fig.~\ref{Fig: 3}(d) we compare the \Eg\, mode broadening (above \Ts) and the \Bx/\By\, splitting (below \Ts) from the two different surfaces. Our results from the cut sample indicate that the structural phase transition changes into a crossover spanning tens of Kelvin above the nominal \Ts, which is consistent with other measurements that report nematicity onsets above \Ts~\cite{Fisher_Science2010,Fisher_Review2011,YiM_PNAS2011,LuDai_Science2014}. 
However, the temperature evolution of the order parameter for the second order phase transition in the polished sample, for which the introduction of strain or stress has been minimized, shows absence of nematic distortion above Ts.

In summary, we reported a Raman scattering study of the in-plane lattice dynamics of EuFe$_2$As$_2$ with two different treatments of the sample side  surface ($ac$): the razor blade cut surface that induces residual stress and the fine polished surface for which the internal strain field is minimized. We observed that the energy splitting of the Fe-As in-plane phonon and the phonon energies from both surfaces are consistent for the whole temperature range. 
The splitting energy scales linearly with the in-plane lattice order parameter of the structural phase transition. However, while our measurements of the strain-free sample indicate that the $C_4$ symmetry breaking occurs only at \Ts\, upon cooling, our results show that the strain field induced by cutting samples with a razor blade breaks the $C_4$ symmetry above \Ts, which may provide an explanation for the observed anisotropy above \Ts\, in various measurements of  samples under uniaxial strain.

W.-L.Z. acknowledges support from NSF (Grant No. DMR- 1104884) and from ICAM (Institute for Complex Adaptive Matter) (NSF-IMI Grant No. DMR-0844115). P.R. and H.D. acknowledge MoST (Grants No. 2011CBA001001 and No. 2015CB921301) and National Natural Science Foundation of China (Grant No. 11274362) of China. A.S.S. and G.B. acknowledge the US Department of Energy, Basic Energy Sciences, and Division of Materials Sciences and Engineering under Awards to ORNL and Grant No. DE-SC0005463 correspondingly.

%


\end{document}